\documentclass[preprint,12pt]{elsarticle}

\usepackage{amssymb}
 \usepackage{lineno}

\journal{Chemical Physics Letters}

\begin{document}

\begin{frontmatter}



\title{Measurement of the Cotton Mouton effect of water vapour}


\author[a1]{F. Della Valle}
\author[a3]{A. Ejlli}
\author[a4]{U. Gastaldi}
\author[a3]{G. Messineo}
\author[a1]{E.~Milotti}
\author[a4]{R.~Pengo}
\author[a3]{L. Piemontese}
\author[a4]{G. Ruoso\corref{cor1}}\cortext[cor1]{Corresponding author} \ead{ruoso@lnl.infn.it}
\author[a3]{G. Zavattini}

\address[a1]{Dip. di Fisica and INFN - Sez. di Trieste, Via A. Valerio, 2 - I - 34127 Trieste (Italy)}
\address[a3]{Dip. di Fisica and INFN - Sez. di Ferrara, Via G. Saragat, 1 - I - 44100 Ferrara (Italy)}
\address[a4]{INFN - Laboratori Nazionali di Legnaro, Viale dell'Universit\`a, 2 - I - 35020 Legnaro (Italy)}



\begin{abstract}
In this paper we report on a measurement of the Cotton Mouton effect of water vapour. Measurement performed at room temperature ($T=301$ K) with a  wavelength
of 1064 nm gave  the value $\Delta n_u = (6.67 \pm 0.45) \cdot 10^{-15}$  for the unit magnetic birefringence (1 T magnetic field and atmospheric pressure).

\end{abstract}



\end{frontmatter}


\section{Introduction}

The Cotton Mouton Effect (CME) refers to the small birefringence that arises in gases in the presence of a magnetic field \cite{cme,cme1}. 
A modern theory of the subject was given by Buckingham and Pople in 1956 \cite{bp} and has been reviewed by Rizzo {\it et al.} \cite{rizzoc}.
CME effect represents a test bench for atomic calculations, especially in the case of low $Z$ gases \cite{Coriani2012}. A determination of the CME effect for a wide variety of gases is crucial in experiments trying to measure the magnetic birefringence of the vacuum \cite{bakalovqso,njp2013,RizzoBMV}. 
In these apparatuses, in fact, 
the CME determines, for each gas species, the maximum tolerable partial pressure beyond which the gas birefringence is no longer negligible with respect to the sought for vacuum effect.
One such species, dominant  in our chamber which is unbakable, is water in its gas form (vapour). 
Up to now, an experimental determination of the CME of water vapour was still missing, and only a theoretical prediction had been given \cite{Ruud1997}. 
The CME of liquid water has been measured in Ref. \cite{Williams}, but the theoretical predictions,  also given in Ref. \cite{Ruud1997}, differ for almost an order of magnitude. Moreover, the sign of the effect is opposite between experiment and theory. Theoretical calculations for the gas phase should be more reliable, which means that an experimental measurement of the CME of water in the gas phase would be very interesting, since it could confirm the sign difference between the two phases. A change of sign from the gas phase to the liquid phase has been experimentally observed for the hyperpolarizability of water \cite{Hyper1, Hyper2}.

In this paper we present a measurement of the CME of water obtained at room temperature and using a laser source with $\lambda$ = 1064 nm. This measurement has been performed using the apparatus of the PVLAS experiment~\cite{njp2013}, whose aim is to detect the magnetic birefringence of the vacuum by means of an ellipsometer coupled to a system of two rotating permanent magnets. The ellipsometer has a very high sensitivity thanks to the use of an ultra high finesse Fabry Perot cavity and of heterodyne detection.

\label{}

\section{Experimental apparatus and method}

A new layout of the PVLAS experiment \cite{njp2013} has been recently designed and built from scratch: as in previous versions, the heterodyne technique is employed, 
 but the magnetic field is now provided by permanent magnets. The measurement principle has been described in several papers (see Ref. \cite{njp2013} for a recent review), we recall here only the specific  characteristics of the new set-up.   

\label{}
\begin{figure}[htb]
\includegraphics[width=\linewidth]{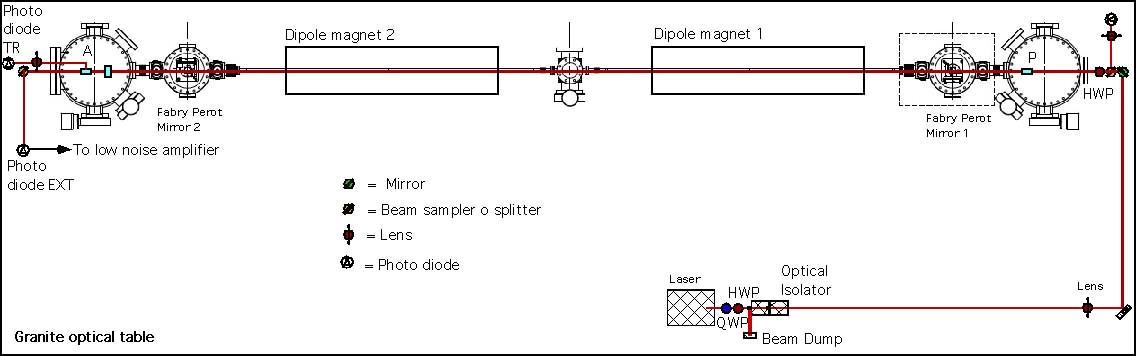}
\caption{Scheme of the apparatus. The granite optical table, dimensions 4.8 x 1.2 m$^2$, is shown together with all the optical components and the five vacuum chambers. The position of the two magnets is also shown. HWP = Half wave plate; P = Polarizer; A = Analyzer;  QWP = Quarter wave plate; TR = transmission; EXT = extinction.}\label{apparatus}
\end{figure}

In figure \ref{apparatus} a scheme of the apparatus is shown.  
A 2 W Nd:YAG NPRO laser is coupled to a high finesse Fabry Perot cavity by  a matching lens and two steering mirrors. The laser output  is elliptically polarized, a quarter wave plate and a half plate are used to maximize the coupling into a two stage optical isolator which prevents back reflected light from re-entering into the laser source. The half wave plate also regulates the amount of laser power used in the experiment. 
The ellipsometer is kept in vacuum: the beam enters the chamber through an optical window with anti reflective coating. Before the window, another half wave plate aligns the polarization along the desired axis.  The ellipsometer is formed by a pair of crossed polarizers: the first one
determines the polarization axis, while the second 
analyzes the beam.
Between the two polarizers, a very high finesse Fabry Perot resonant cavity is used as an optical path amplifier; the laser is frequency locked to the cavity \cite{rsi}. A polarization modulator (PEM) placed before the analyzer is used for heterodyne detection. The two last elements are decisive in obtaining extremely high sensitivity for detecting birefringences.

Magnetic birefringence is generated by two permanent dipole magnets that rotate independently around the light propagation direction in order to modulate the effect. 
The magnets rotation is given, through a system of gears and  belts, by two synchronous electric motors whose frequency is controlled by  two stabilized and phase locked frequency generators.
The optical and vacuum components of the apparatus are mounted on an air suspended granite optical table. A computer controlled feedback system, acting on the table legs, provides seismic isolation. In order to avoid vibrations of the ellipsometer generated by the magnets rotation, the magnets  are hanging from an aluminium structure mechanically decoupled from the rest of the optical table.

The extincted beam exiting the analyzer is collected onto an InGaAs photodiode, and the resulting photocurrent is then amplified using a low noise current amplifier. The output is frequency down-converted using a lock-in amplifier having  the driving signal of the polarization modulator as a reference oscillator. A FFT analyzer computes the frequency spectrum of the lock-in output, both in amplitude and phase. 
A magnetic field probe continuously monitors the magnetic field component along one direction. This signal is used as a trigger by the spectrum analyzer, thus ensuring  that the spectrum of the lock-in signal is always referenced to the same magnet angular position.

The vacuum structure comprises five UHV metal chambers plus a 16 mm internal bore pyrex pipe through each of the two magnets. The chambers are made of non magnetic steel and titanium, in order to avoid coupling effects with the magnets' stray field. 
Vacuum pumping systems, each composed of  a scroll pump and a turbo molecular pump are used to evacuate the chambers down to $\simeq 10^{-7}$ mbar. By using a leak valve it is possible to insert into the chambers an amount of gas whose pressure is measured with  a capacitive transducer.

The signal we are looking for is an ellipticity $\psi$ induced on the beam by the CME. This is related to the anisotropy of the index of refraction $\Delta n = n_\| - n_\bot$, where the $\| $ and $\bot$ axes are defined with respect to the magnetic field direction.
Following Ref. \cite{rizzoc}, we define a birefringence for unit field (1 T) and  atmospheric pressure $p_{\rm atm}$ as:
\begin{equation}
\Delta n_u= \Delta n \left( \frac{1}{ B [ {\rm T} ]} \right)^2\left(\frac{p_{\rm atm}}{p}\right)\label{delta}
\end{equation}
 where $p$ is the pressure and $B$ the magnetic field amplitude. For a dipole magnet of length $L$:

\begin{equation}
\psi(t)= 2 \frac{F}{\lambda} \sin 2 \theta(t) \left( \int_0^L B^2 [ {\rm T} ] \,\,dz \right) \Delta n_u \frac{p}{p_{\rm atm}}
\label{deltanu}
\end{equation}
where we have already taken into account the $2 F/ \pi$ amplification factor resulting from a cavity with finesse $F$ (see Ref. \cite{rizzoc}). The angle $\theta$ is measured from the polarization of the light to the direction of the magnetic field $B$, which lies in the plane perpendicular to the light propagation direction (dipole field).
The low frequency lock-in output has a spectral component at twice the magnet rotation frequency, whose amplitude is a measure of $\psi$. Since the data acquisition starting time is always given at the same angular position of the rotating magnet, phase information can also be extracted and used  to measure the relative sign of the ellipticity. The magnet rotation frequency was 4 Hz, thus the resulting signal frequency is 8 Hz.
\begin{figure}[h]
\includegraphics[width=\linewidth]{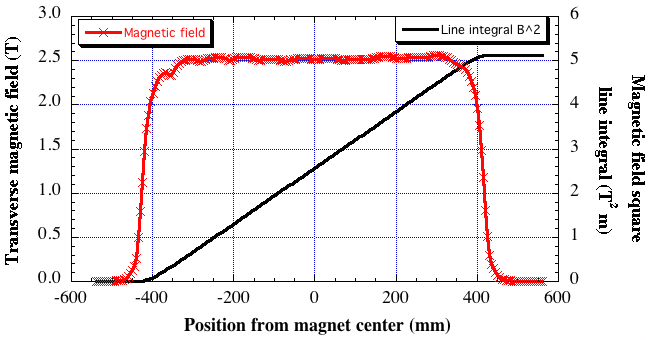}
\caption{Measurement of the magnetic field of the permanent magnet. }
\label{Magnet}
\end{figure}
For the measurement presented here only one magnet has been used. Its magnetic field has been  measured using a three-axis Hall probe: the result, showed in figure \ref{Magnet}, corresponds to $5.12\pm0.04$ T$^2$ m for the field square line integral.

Measurements have been performed with water vapour and nitrogen. The latter has been used to test the apparatus and gives an absolute reference for the sign of the CME: the magnetic birefringence induced on nitrogen is in fact negative \cite{rizzoc}.
We used spectroscopic type nitrogen, and the chamber and the gas line were evacuated to high vacuum before being filled with the gas. 
Water vapour is prepared by repeated cycles of vacuum pumping with a turbo pump above the free surface of the water contained in a glass ampoule. After the first couple of cycles the water will no longer boil. After pumping, the  water vapour in equilibrium at about 20 ~mbar with the liquid is then allowed to leak into the experimental chamber.

For each measurement we measured the finesse $F$ of the cavity from the cavity decay time $\tau$:

\begin{equation}
\tau=\frac{Fd}{\pi c}
\end{equation}
where $c$ is the speed of light and $d=3.303 \pm 0.005$ m is the cavity length. 
We measured $\tau=330\pm20$ $\mu$s, corresponding to a finesse $F=94200\pm5700$.
Since we noticed some instabilities in the cavity when using nitrogen which depended on laser power, the amount of light used  was different for the two gases:  the circulating power in the resonant cavity was about 60 W for nitrogen and about 270 W for water vapour. 

The temperature of the system was monitored with a commercial sensor. The average value for all the data sets is $301\pm 1$ K. 
\section{Results}

We took one data set with nitrogen and two data sets with water vapour. 
For each data set the ellipsometer conditions were kept fixed and the pressure of the gas sample was changed in steps. 
The two data sets with water differ only in the direction of   pressure variation: in one set (WP1) pressure was raised from vacuum to a maximum value.
In the other set (WP2) the water vapour pressure was decreased from a maximum to a minimum
by pumping the chamber.

\begin{figure}[htb]
\includegraphics[width=\linewidth]{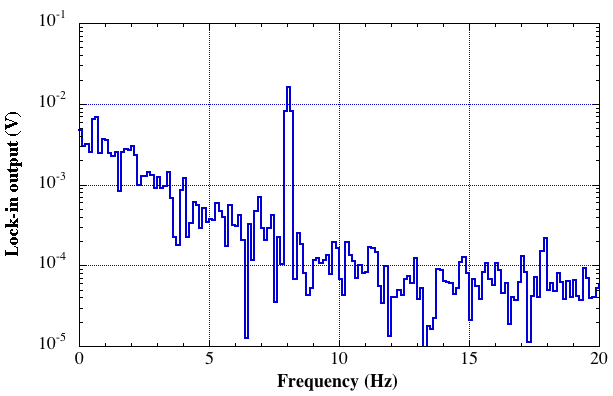}
\caption{Frequency spectrum of the lock-in output showing the peak corresponding to an ellipticity of $2.5\cdot10^{-5}$ due to 4 mbar of water vapour in the chamber. The peak frequency is 8 Hz since the magnet rotation frequency is 4 Hz.}\label{spettro}
\end{figure}

For each pressure $p_i$ we calculated the ellipticity $\psi_i$, with its phase $\varphi_i$, from the FFT spectrum. 
The statistical error on $\psi_i$ is obtained by adding quadratically the uncertainties on the cavity transmitted power, the photo-elastic modulator signal amplitude  and the noise floor  around 8 Hz on the FFT spectrum.
 The integration time
 was about 100 s for each pressure value.
Figure \ref{spettro} shows the frequency spectrum obtained with 4 mbar of  water vapour in the chamber. The measured ellipticity is $\psi= 2.5\cdot10^{-5}$, with a signal to noise ratio of order $\simeq 10^2$. 
\begin{figure}[htb]
\includegraphics[width=\linewidth]{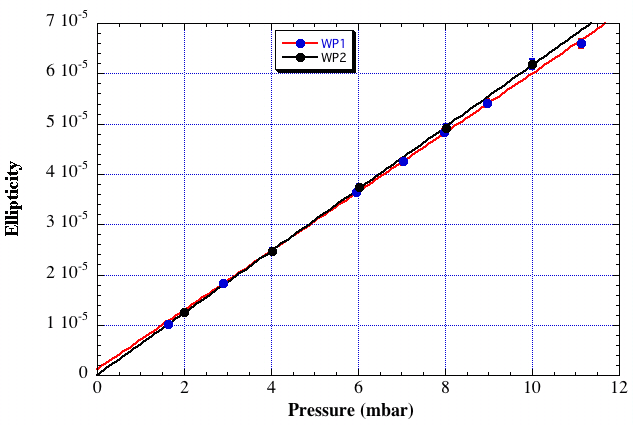}
\caption{Measured ellipticity $\psi$ as a function of water vapour pressure $p$ for the two data sets WP1 and WP2.  Experimental points with error and fit with a linear function of the type $\psi(p) = a + b p$ are shown. }\label{water}
\end{figure}

Figure \ref{water} shows the values of ellipticity obtained for the water measurements WP1 and WP2. The experimental points are fitted with  straight lines of the type $\psi(p) = a + b p$. The intercept $a$ is an artifact not related with the presence of gas in the apparatus. The slope $b$ is directly connected to the unit birefringence defined in equation (\ref{deltanu}).

\begin{table}[h]
\caption{Summary of the measurements of the CME for nitrogen and water vapour. Parameters $a$ and $b$ describe the linear fit to the measured data. The 'avg. phase' is the average phase at the signal frequency of the spectrum of the lock-in output. The $\chi^2/{\rm d.o.f.}$ is the reduced chi square value for the fit. }
\begin{center}
\begin{tabular}{|c|c|c|c|c|}
\hline
 Gas &  $a$ & $b$ [1/mbar] & avg. phase & $\chi^2/{\rm d.o.f.}$ \\
\hline
N$_2$ & $(2.8\pm4.5)10^{-5}$ & $(2.31\pm0.07)10^{-4}$ & $-97.0\pm0.6$& 1.1 \\
\hline
H$_2$O (WP1)&   $(12.1\pm3.5)10^{-7}$ & $(5.88\pm0.07)10^{-6}$ & $82.7 \pm 0.3$& 0.6 \\
\hline
H$_2$O (WP2)&  $(0.7\pm4.7)10^{-7}$ & $(6.16\pm0.10)10^{-6}$ & $83.3 \pm 0.4$& 0.22\\
\hline
\end{tabular}
\end{center}
\label{risulta}
\end{table}%

Table \ref{risulta} shows all the results. The average phases of all experimental points have also been included. As it can be seen, the sign of the CME of water vapour is opposite to that of nitrogen, i.e. it is positive. 
The final value of the unit birefringence has been obtained by reversing equation (\ref{deltanu}):
\begin{equation}
\Delta n_u = b \frac{\lambda p_{\rm atm}}{\left( \int_0^L B^2 [ {\rm T} ] \,\,dz \right) F}.
\end{equation}
For the case of water we take the weighted average of the two values of the parameter $b$. In this case we use the half difference of the two values as an estimate of the overall uncertainty.  The uncertainties for each set of measurements result from the statistical error on the $\psi_i$'s and from the measurements  of the magnetic field, of the finesse and of the quantum efficiency of the photodetector. The final values are: 
\begin{eqnarray}
\Delta n_u ({\rm nitrogen})& =& (-2.6 \pm 0.2) \cdot 10^{-13}\\ 
\Delta n_u ({\rm water}\,{\rm vapour})& =& (6.67 \pm 0.45) \cdot 10^{-15}
\end{eqnarray}
where we have assumed that the CME for nitrogen is negative. The result for nitrogen is compatible with previous measurements \cite{rizzoc}.



\section{Discussion}

The first estimate of the CME for water in the gas phase has been given in Ref.  \cite{Rizzo1995}. The value presented there was then corrected in a subsequent paper \cite{Ruud1997}, where also new estimates have been given. Theoretical calculations compute the so called molar Cotton Mouton constant $_mC$ (expressed in $\rm{cm}^3\, \rm{G}^{-2}\, \rm{mol}^{-1}$), which is connected to the unit birefringence by the relation \cite{rizzoc}:

\begin{equation}
\Delta n_u=\frac{1.64518\cdot10^7}{T}\,\, _mC\,\, 
\label{conv}
\end{equation}
where $T$ is the absolute temperature.
In table \ref{TabellaB},
a comparison between our result and the theoretical calculation is given. 


\begin{table}[htdp]
\caption{Summary of theoretical and experimental values for the CME of water in the gas phase. The experimental value has been computed  using formula (\ref{conv}) at the temperature of $T=301$ K. The value (a) is the value of Ref.  \cite{Rizzo1995} as corrected in Ref. \cite{Ruud1997}. The values (b) and (c) are taken from Ref. \cite{Ruud1997}. MCSCF = Multiconfigurational self-consistent field; SCF = Self - consistent field; CASSCF =  Complete active space self-consistent field. The units for $_mC$ are $\rm{cm}^3\, \rm{G}^{-2}\, \rm{mol}^{-1}$.}
\begin{center}
\begin{tabular}{|c|c|}
\hline
Theory & $_mC$ (283.15 - 293.15 K) \\
\hline
MCSCF (a) & $10.4 \cdot 10^{-20}$   \\
\hline
SCF (b) & $9.67 \cdot 10^{-20}$ \\
\hline
CASSCF (c) & $10.13 \cdot 10^{-20}$ \\
\hline
& \\
\hline
  Experiment & $_mC$ ($301$ K)\\
\hline
& $(12.2 \pm 0.8) \cdot 10^{-20}$\\
\hline
\end{tabular}
\end{center}
\label{TabellaB}
\end{table}%

The most recent experimental result for liquid water \cite{Williams} gives $_mC = (- 1.18 \pm 0.15) \cdot 10^{-18}\,\,\rm{cm}^3\, \rm{G}^{-2}\, \rm{mol}^{-1}$ (in the interval  $T= 283.15-293.15$~K). Our results confirms the  sign difference of the CME between the liquid and the gas phase. Moreover, while theoretical calculations in the liquid phase are at least a factor 8 smaller than the experimental measurement \cite{Ruud1997}, this discrepancy is actually recovered in the case of the gas phase, where the difference is smaller than 3 $\sigma$.  

\section{Conclusions}

We have measured for the first time the Cotton Mouton effect for water in the gas phase. This measurement shows evidence of a sign difference for the effect in liquid water with respect to water vapour.
The measured value of the effect differs from the theoretical calculations, but in a much less dramatic way with respect to the case of liquid water.

Finally, given the measured value of $\Delta n_u$ for water vapour, the partial pressure of water vapour which will give the same birefringence as vacuum is $\approx 6 \cdot 10^{-7}$ mbar.










\end{document}